\documentclass[prl,twocolumn,showpacs,amsmath,amssymb]{revtex4}
\usepackage{graphicx}
\usepackage{dcolumn}
\usepackage{bm}

\begin{document}

\title{Dipole responses in Nd and Sm isotopes with shape transitions
}

\author{Kenichi Yoshida and Takashi Nakatsukasa}
\affiliation{RIKEN Nishina Center for Accelerator-Based Science, Wako, Saitama 351-0198, Japan
}%

\date{\today}

\begin{abstract}
Photoabsorption cross sections of Nd and Sm isotopes 
from spherical to deformed even nuclei are systematically
investigated by means of the quasiparticle-random-phase approximation 
based on the Hartree-Fock-Bogoliubov ground states (HFB+QRPA)
using the Skyrme energy density functional.
The gradual onset of deformation in the ground states
as increasing the neutron number
leads to characteristic features of the shape phase transition. 
The calculation well reproduce the isotopic dependence of 
broadening and emergence of a double-peak structure in the cross sections 
without any adjustable parameter.
We also find that the deformation plays a significant role for
low-energy dipole strengths.
The $E1$ strengths are fragmented and considerably lowered in energy.
The summed $E1$ strength up to 10 MeV is enhanced 
by a factor of five or more.
\end{abstract}

\maketitle

Density functional theory has been widely used to describe 
a variety of quantum many-body systems~\cite{par83} including
nuclear many-body systems~\cite{ben03}.
Recent advances in the computing capability
together with the highly-developed techniques in
the nuclear energy-density-functional (EDF) method 
allow us to calculate
the ground-state properties of nuclei in the entire mass region~\cite{sto03}.
The nuclear ground-state deformation is one of them, which is an example
of the spontaneous breaking of rotational symmetry.
Experimental evidences of the nuclear shape changes are related to
low-lying quadrupole collectivity, such as
the ratio of the excitation energy of $2^{+}$ and $4^{+}$ states $E_{4+}/E_{2+}$,
the reduced transition probability $B(E2;2^{+}\rightarrow 0^{+})$, etc.
However, it is known that
the nuclear deformation also affects the high-frequency collective modes
of excitation, giant resonances (GRs)~\cite{BM2,har01}. 
For instance, the peak splitting of the giant dipole resonance (GDR),
which is caused by the different frequencies of oscillation
along the long and short axes,
has been observed in experiments~\cite{BF75}.
A typical example of the shape phase transition form spherical to
deformed ground states and an evolution of the deformation splitting in GDRs
have been observed in Nd and Sm isotopes~\cite{BM2,har01,BF75,car71,car74}.
In this letter, we report a first systematic calculation of
electric dipole ($E1$) responses for these heavy isotopes
with the shape phase transition,
using a non-empirical approach with the Skyrme EDF, namely,
the quasiparticle-random-phase-approximation based on the
Hartree-Fock-Bogoliubov ground states (HFB+QRPA).

GDRs in heavy deformed systems have been previously investigated
using the separable QRPA with Skyrme EDF~\cite{nes06,nes08}. 
The separable approximation in the QRPA, perhaps, provides a good approximation
for the GDR. 
It is, however, difficult to analyze the low-lying states
because the structure of normal modes is non-trivial and
significantly affected by the detailed shell structure.
The low-energy $E1$ strengths,
which are often discussed as the pygmy dipole resonance (PDR), 
have attracted a considerable interests.
A nonstatistical distribution of the $E1$ strengths close to the threshold has 
a strong impact to the astrophysical r-process nucleosynthesis~\cite{gor02}. 
In addition, 
the PDR is a typical example of exotic collective modes
expected in neutron-rich nuclei and has been extensively
studied with the EDF approaches~\cite{paa07}. 
However, the role of deformation on the PDR has been studied only for
light nuclei \cite{yos09b,eba10},
except for a recent study on Sn isotopes
with the relativistic EDF~\cite{pen09}.

We have developed a new parallelized computer code of the HFB+QRPA,
which is an extended version of that developed in Ref. \cite{yos08},
to add the residual spin-orbit interaction.
We expect that the residual Coulomb plays only a minor role~\cite{ter05,eba10}. 
To describe the nuclear deformation 
and the pairing correlations, simultaneously, in good account of the continuum,
we solve the HFB equations~\cite{dob84}
in the coordinate space using cylindrical coordinates
$\boldsymbol{r}=(\rho,z,\phi)$ with a mesh size of
$\Delta\rho=\Delta z=0.6$ fm and a box
boundary condition at $(\rho_{\mathrm{max}},z_{\mathrm{max}})=(14.7, 14.4)$ fm.
We assume axial and reflection symmetries in the ground state.
The quasiparticle (qp) states are truncated according to the qp
energy cutoff at $E_\alpha \leq 60$ MeV. 
We introduce the additional truncation for the QRPA calculation,
in terms of the two-quasiparticle (2qp) energy as
$E_{\alpha}+E_{\beta} \leq 60$ MeV.
This reduces the number of 2qp states to, for instance,
about 38 000 for the $K^{\pi}=0^{-}$ excitation in $^{154}$Sm.
To calculate the QRPA matrix elements and to diagonalize the matrix, 
it takes about 390 CPU hours and 135 CPU hours, respectively.  
A detailed description of the method and its computational aspects
will be soon published elsewhere. 
The similar calculations of the HFB+QRPA for axially deformed nuclei  
have been recently reported~\cite{per08,pen09,los10,ter10}.

For a normal (particle-hole) part of the EDF,
we employ the SkM* functional~\cite{bar82}. 
For the pairing energy, we adopted the one in Ref. \cite{yam09}
that depends on both
the isoscalar and isovector densities, 
in addition to the pairing density, with the parameters given in
Table~III of Ref.~\cite{yam09}.

Since the full self-consistency between the static mean-field
calculation and the dynamical calculation is slightly broken by the
neglected residual terms and the truncation of the 2qp states,
the spurious states may appear at finite excitation energies. 
In the present calculation, the excitation energies of the spurious states 
with $K^{\pi}=0^{-}$ and $1^{-}$,
corresponding to the center-of-mass motion,
become imaginary in $^{154}$Sm, 1.46$i$ MeV and 1.60$i$ MeV, respectively.
Small contamination of the spurious component does not affect the GDRs
because they are far apart in energy.
The $E1$ strength of low-lying dipole states might be slightly
influenced.
However, it will not affect the conclusion of the present study.

\begin{figure}[t]
\begin{center}
\includegraphics[scale=0.8]{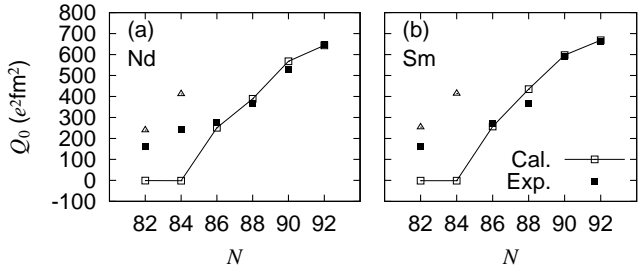}
\caption{Intrinsic electric quadrupole moments of Nd and Sm isotopes. 
For spherical nuclei, values extracted from calculated $B(E2)$ are shown by
triangles.
}
\label{deformation}
\end{center}
\end{figure}

Figure~\ref{deformation} shows the ground-state deformation of Nd and Sm
isotopes obtained with the HFB calculation.
The calculated intrinsic electric quadrupole moments are compared 
with the experimental values~\cite{ENSDF}. 
The calculation well reproduces the evolution of quadrupole deformation
for $N \geq 86$.
For spherical nuclei with $N=82$ and 84,
we also plot the values deduced from
the $B(E2;0^{+}\rightarrow 2^{+})$ obtained by the QRPA calculation. 
Collectivity of the $2^{+}$ state is apparently overestimated at $N=84$,
because these nuclei are so soft with respect to the quadrupole deformation
that the QRPA cannot describe the $2^{+}$ state properly. 

Based on these HFB ground states, we perform the QRPA calculation
to obtain the excitation energies, $\hbar\omega_i$,
and the transition matrix elements, $\langle i|\hat{F}^{1}_{1K}|0 \rangle$.
The photoabsorption cross section is calculated as 
\begin{align}
\sigma_{\mathrm{abs}}(E)&=\dfrac{4\pi^{2}E}{\hbar c}\sum_{K=-1}^{1}\dfrac{dB(E,F^{1}_{1K})}{dE}, \\
\dfrac{dB(E,F^{1}_{1K})}{dE} &=\dfrac{2E\gamma}{\pi}\sum_{i}
\dfrac{\tilde{E}_{i} |\langle i|\hat{F}^{1}_{1K}|0 \rangle|^{2}}{(E^{2}-\tilde{E}_{i}^{2})^{2}+E^{2}\gamma^{2}},
\end{align}
where $\tilde{E}_{i}^{2}=(\hbar \omega_{i})^{2}+\gamma^{2}/4$~\cite{BM2}. 
The smearing width $\gamma$ is set to 2 MeV, which is supposed to simulate
the spreading effect, $\Gamma^\downarrow$, missing in the QRPA.
The electric dipole operator $\hat{F}^{1}_{1K}$ is defined 
as Eq.~(6-175) of Ref.~\cite{BM2}.

\begin{figure}[t]
\begin{center}
\includegraphics[scale=0.66]{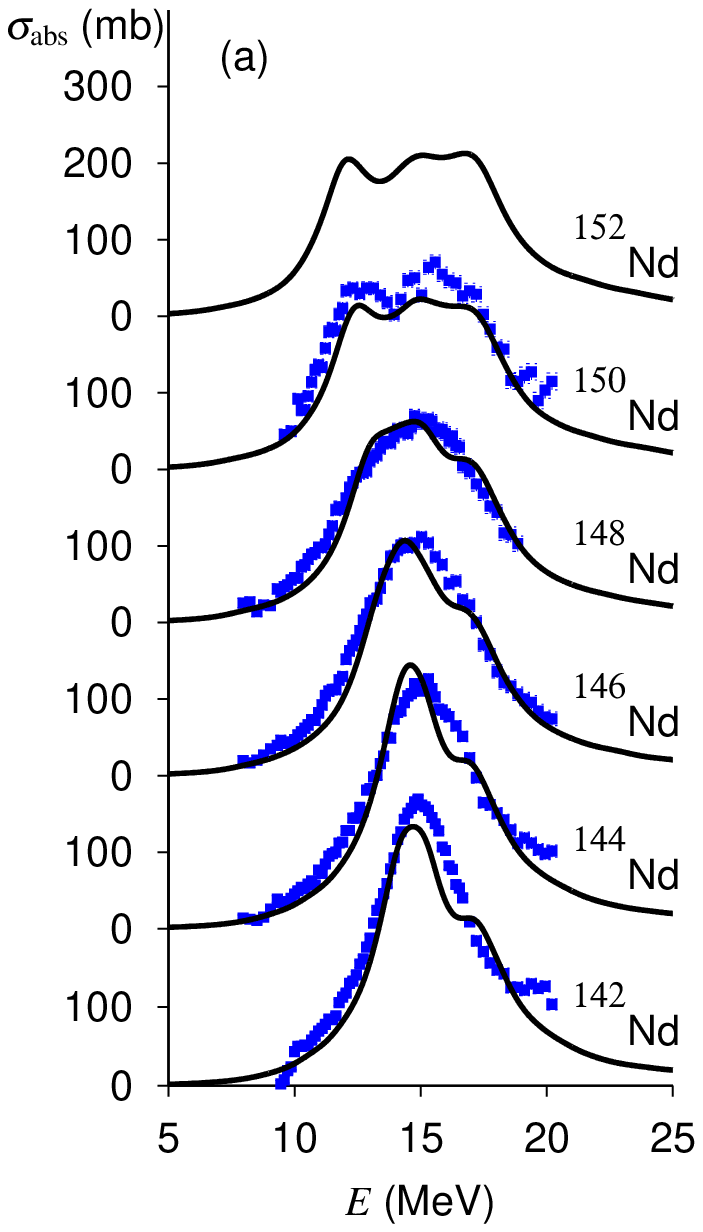}
\includegraphics[scale=0.66]{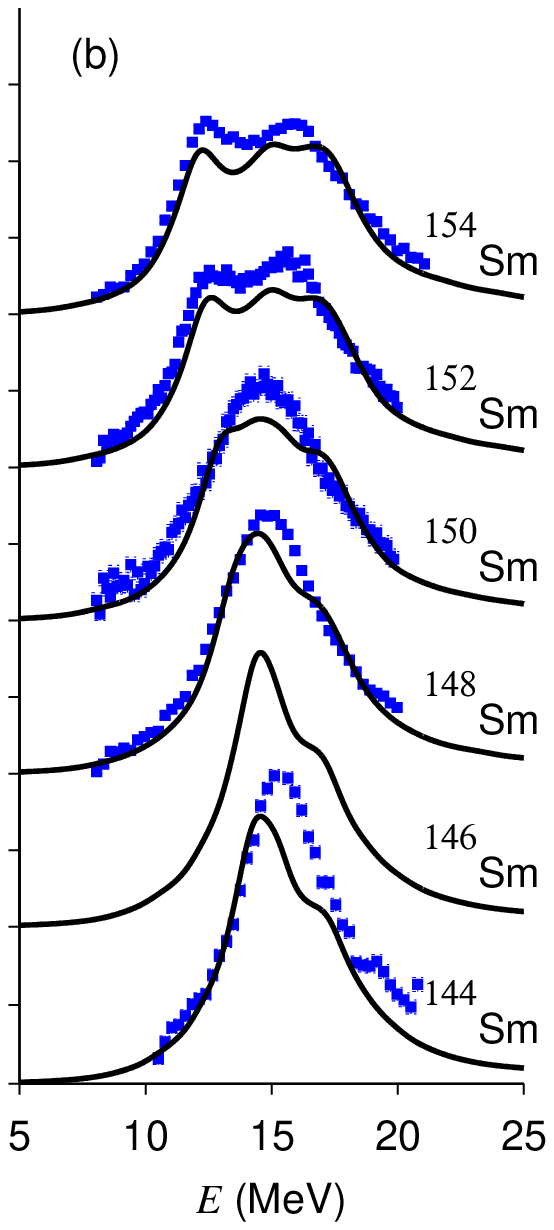}
\caption{(Color online) 
Photoabsorption cross sections in Nd and Sm isotopes as functions of
photon energy.
The experimental data~\cite{car71,car74} are denoted by filled squares.
}
\label{photo}
\end{center}
\end{figure}

We show in Fig.~\ref{photo} the calculated photoabsorption cross sections 
in the GDR energy region 
together with the available experimental data~\cite{car71,car74}. 
The GDR peak energies well agree with experimental values, and
produces the deformation splitting in $^{150,152}$Nd and $^{152,154}$Sm.
Previously, the separable QRPA calculations with SkM*~\cite{nes06,nes08} 
produce a significantly larger second peak for these deformed nuclei,
which disagrees with the experiment.
This is diminished in the present full QRPA calculation.
The GDR width calculated with $\gamma=2$ MeV
is also in good accordance with the experimental values.
The QRPA accounts for the Landau damping, which is a fragmentation of the
GDR strength into nearby 2qp states,
but not for the spreading effect which corresponds to
a fragmentation into more complex states.
The nice agreement on the broadening indicates that 
the smearing width $\gamma=2$ MeV
has a good correspondence with the spreading width
$\Gamma^{\downarrow}$ in these nuclei. 

The isotopic dependence of the peak broadening
is well reproduced,
surprisingly, even for the transitional nuclei.
The width for $N=82$ and 84 is calculated as $\Gamma\approx 4.5$ MeV,
and it gradually increases to about 6 MeV for $N=88$, then
the splitting becomes visible for $N\geq 90$. 
Here, the width $\Gamma$ is evaluated by fitting the calculated cross section 
with a Lorentz line. 
This broadening effect
is commonly interpreted as the mode-mode coupling effects to the low-lying
collective modes~\cite{har01}.
In the present QRPA calculation,
the mode coupling is not explicitly taken into account. 
However, the QRPA based on the deformed HFB state may implicitly
include a part of the coupling effect.
Figure~\ref{photo} shows that the isotopic dependence
can be well accounted for
by the gradual increase of the ground-state deformation.
However, a small increase of the width
from $^{142}$Nd to $^{144}$Nd observed in the experiment
cannot be fully reproduced in the calculation.
Since the HFB calculation produces the spherical ground state for $^{144}$Nd,
this requires an explicit higher-order calculation beyond the QRPA.

We may notice another small disagreement in the peak shape:
The calculated GDR peak has a shoulder in the spherical nuclei, 
and this shoulder is becoming the third peak in the deformed nuclei.
This is due to the Landau fragmentation,
however, this feature is not clearly observed in the experiments.
As is discussed in the followings,
detailed properties of the Landau fragmentation
depend on the choice of the Skyrme EDF.
For instance,
the fragmentation effect becomes weaker with
the SkP functional, to give a better agreement with the experiments.

\begin{figure}[t]
\begin{center}
\includegraphics[scale=0.65]{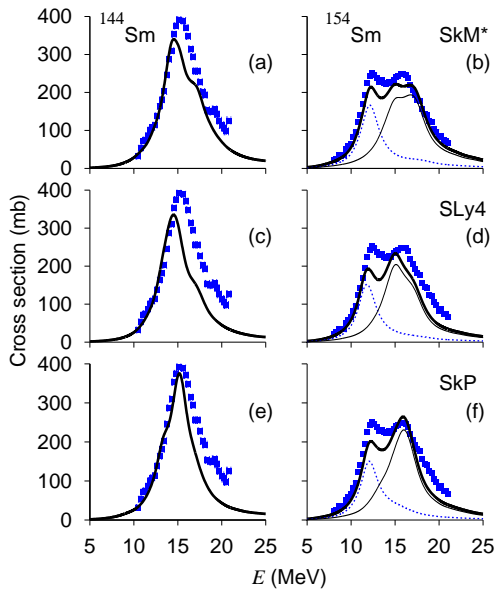}
\caption{(Color online)
Photoabsorption cross sections of $^{144}$Sm and $^{154}$Sm employing the SkM*, 
SLy4 and SkP functionals. The $K^{\pi}=0^{-}$ and $K^{\pi}=1^{-}$ components are 
shown by dotted and thin solid lines, respectively in $^{154}$Sm.
}
\label{144_154Sm}
\end{center}
\end{figure}

Figure~\ref{144_154Sm} shows photoabsorption cross sections in $^{144,154}$Sm
obtained by employing the 
SLy4~\cite{cha97} and SkP~\cite{dob84}.
The experimental GDR peak structure is reproduced 
not only by using the SkM* functional but also
by using other commonly-used Skyrme functionals,
though the SLy4 gives slightly smaller GDR peak energies. 
The SkP reproduces the energy and the shape best among the three functionals.
For spherical nuclei, the cross section obtained with the SkP functional
can be nicely fitted by a single Lorentzian curve. 
For deformed nuclei, the lower ($K=0$) peak shows a Lorentz shape for
any of the functional, while the upper ($K=1$) peak shows visible distortion
for SkM* and SLy4.
This difference may be due to different properties of the Landau
fragmentation because the 2qp states in the background are more widely
spread in energy with a smaller effective mass.
Actually, the effective mass is largest ($m^*/m\approx 1$) in SkP.

\begin{figure}[t]
\begin{center}
\includegraphics[scale=0.64]{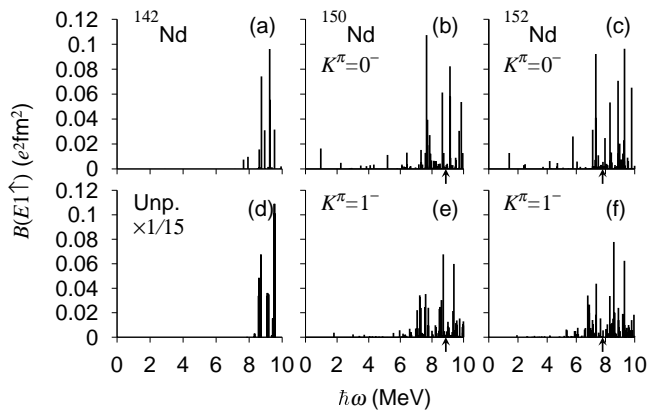}
\caption{Calculated low-energy $E1$ strengths
$B(E1;0^{+}\rightarrow 1^{-})$ as functions of energy 
in $^{142,150,152}$Nd. 
The arrow indicates the neutron emission threshold energy.
The panel (d) shows the unperturbed strengths multiplied by 1/15 in $^{142}$Nd. 
The neutron threshold of $^{142}$Nd is 11.6 MeV.
}
\label{Nd_dipole_strength}
\end{center}
\end{figure}

\begin{figure}[t]
\begin{center}
\includegraphics[scale=0.64]{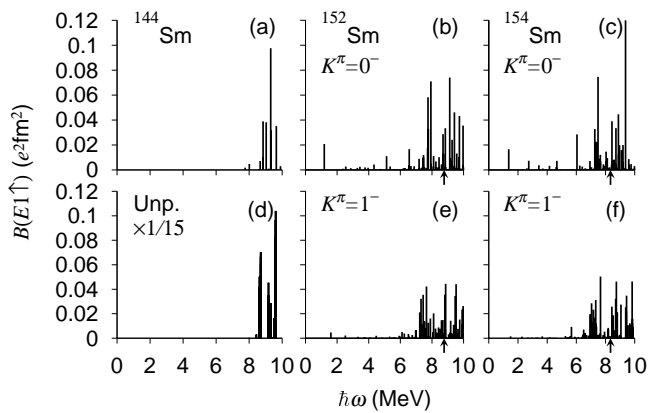}
\caption{Same as Fig.~\ref{Nd_dipole_strength} but in $^{144,152,154}$Sm.
The panel (d) shows the unperturbed strengths multiplied by 1/15 in $^{144}$Sm.
The neutron threshold of $^{144}$Sm is 11.9 MeV.}
\label{Sm_dipole_strength}
\end{center}
\end{figure}

Next, we discuss the low-energy $E1$ strengths. 
Photon scattering experiments
at Technische Universit\"at Darmstadt~\cite{zil02} 
reported the $1^{-}$ states up to 9.9 MeV and found 
the concentration of the dipole strength
in $N=82$ semi-magic nuclei at energies between 5.5 and 8 MeV.
Figures~\ref{Nd_dipole_strength} and \ref{Sm_dipole_strength} show the 
the $E1$ strengths below 10 MeV calculated with the SkM* functional.
The SLy4 and SkP functionals provide very similar results.
In the spherical $^{142}$Nd and $^{144}$Sm nuclei, we can see a 
concentration of the dipole strength in between 8 and 10 MeV. 
The same concentration can be seen in the unperturbed strength distribution, 
which may suggest a weak collectivity. 
Apparently, the calculated strength distribution is too high in energy.
Similar disagreements have been also observed in 
the relativistic QRPA calculation~\cite{paa03}.
However, the $B(E1\uparrow)$ value summed up to 10 MeV in $^{144}$Sm
is $0.27$ $e^{2}$fm$^{2}$ which reasonably agrees with experimental value  
0.20 $e^{2}$fm$^{2}$.
It seems to suggest that these low-energy $E1$ strengths are
redistributed to two-phonon ($2^{+}\otimes 3^{-}$) and multi-phonon states,
which are located at energies $3\sim 7$ MeV,
by higher-order coupling effects \cite{zil02}.
It is quite challenging to describe such multi-phonon states
on the basis of the nuclear EDF method~\cite{sar04,lit10}, 
however, it is beyond the scope of the present analysis.

\begin{figure}[t]
\begin{center}
\includegraphics[scale=0.86]{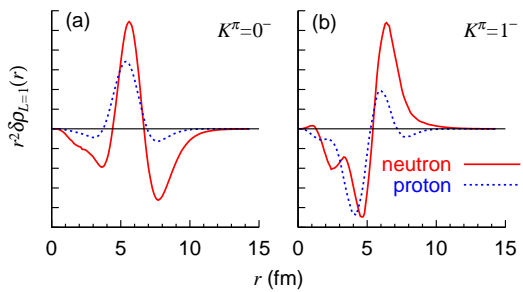}
\caption{(Color online)
Calculated transition densities of the $K^{\pi}=0^{-}$ state at 7.49 MeV and 
of the $K^{\pi}=1^{-}$ state at 7.33 MeV in $^{154}$Sm.
}
\label{154Sm_trans_density}
\end{center}
\end{figure}

In contrast to the spherical nuclei,
the deformed nuclei show significant fragmentation of the $E1$ strength
into low-energy states.
For $^{150,152}$Nd and $^{152,154}$Sm,
the calculated energies of the lowest $K^{\pi}=0^{-}$ states are
0.97, 1.40, 1.10, and 1.37 MeV, 
and those of $K^{\pi}=1^{-}$ states are 
1.80, 1.93, 1.60, and 1.49 MeV, respectively. 
These values agree well with available experimental values of
octupole vibrational states \cite{sood91}.
The deformation significantly increases the total low-energy $E1$ strength.
Actually, the summed strengths up to 10 MeV,
$\sum B(E1\uparrow) \approx 1.5$ $e^2$fm$^2$, are
about five times larger than those of spherical nuclei.
It may be of significant interest to study
how this low-energy $E1$ enhancement due to deformation affects
the element synthesis scenario.

We also confirm that the dipole states around 7 MeV have a character 
different from the GDR.
In Fig.~\ref{154Sm_trans_density}, the transition densities,
approximately projected to the laboratory frame~\cite{PR08} 
are shown in arbitrary scales.
These dipole states predominantly have an isoscalar character: 
The proton and neutron transition densities have the same sign. 
The neutron dominance can be seen at $r > 7$ fm, however, the proton
contribution does not completely vanishes.
These characters of the PDR are consistent with the previous QRPA analysis
for spherical nuclei of $N=82$ isotones \cite{paa03}.
Increasing $N/Z$ ratio away from the stability,
we expect that the neutron-dominant character develops in the surface region.

In summary, we have investigated effects of the shape transition on
$E1$ strength distribution in the rare-earth nuclei,
using the newly developed parallelized HFB+QRPA calculation code 
with the Skyrme EDF.
This enables us to simultaneously study both high-energy GRs and
low-energy collective/non-collective states.
The typical characteristics of the GDR for shape phase transition
from spherical to deformed nuclei, 
especially the isotopic dependence of
broadening and splitting of the GDRs,
are extremely well reproduced in the calculation.
We have also found that the deformation plays a significant role for
the low-energy $E1$ strength distribution:
The $E1$ strength is distributed to low-energy states and the total
strength at $E<10$ MeV is roughly five times enhanced,
compared to the spherical nuclei.
The low-energy strengths in spherical $^{144}$Sm are calculated too high
in energy.
Inclusion of the higher-order mode-mode coupling is desired for further
improvements.
Systematic calculations with the HFB+QRPA
for spherical-to-deformed and light-to-heavy nuclei
help us not only to understand and to predict new types of collective modes of
excitation, but also to shed light on the nuclear EDF of new generations.

One of the authors (K.Y.) is supported by the Special Postdoctoral 
Researcher Program of RIKEN. 
The work is supported by Grant-in-Aid for Scientific Research
(Nos. 21340073 and 20105003) and by the Joint Research Program at
Center for Computational Sciences, University of Tsukuba.
The numerical calculations were performed 
on RIKEN Integrated Cluster of Clusters (RICC).


\begin{thebibliography}{99}
\bibitem{par83}
R.~G.~Parr, Ann. Rev. Phys. Chem. {\bf 34}, 631 (1983).

\bibitem{ben03}
M.~Bender {et al.}, 
Rev. Mod. Phys. 75 (2003) 121.

\bibitem{sto03}
M.~V.~Stoitsov {\it et al.},
Phys. Rev. C {\bf 68}, 054312 (2003).

\bibitem{BM2}
A.~Bohr and B.~R.~Motteleson, 
{\it Nuclear Structure}, vol.~II (Benjamin, 1975; World Scientific, 1998).

\bibitem{har01}
M.~N.~Harakeh and A. van der Wounde,
{\it Giant Resonances}
(Oxford, 2001).

\bibitem{BF75}
B.~L.~Berman and S.~C.~Fultz, 
Rev. Mod. Phys. {\bf 47}, 713 (1975).

\bibitem{car71}
P.~Carlos {\it et al.}, 
Nucl. Phys. {\bf A172}, 437 (1971).

\bibitem{car74}
P.~Carlos {\it et al.}, 
Nucl. Phys. {\bf A225}, 171 (1974).

\bibitem{nes06}
V.~O.~Nesterenko {\it et al.}, 
Phys. Rev. C {\bf 74}, 064306 (2006).

\bibitem{nes08}
V.~O.~Nesterenko {\it et al.}, 
Int. J. Mod. Phys. E {\bf 17}, 89 (2008).

\bibitem{gor02}
S.~Goriely and E.~Khan, Nucl. Phys. {\bf A706}, 217 (2002).

\bibitem{paa07}
N.~Paar {\it et al.},
Rep. Prog. Phys. {\bf 70}, 691 (2007).

\bibitem{yos09b}
K.~Yoshida, Phys. Rev. C {\bf 80}, 044324 (2009).

\bibitem{eba10}
S.~Ebata {\it et al.}, 
arXiv:1007.0785.

\bibitem{pen09}
D.~P.~Arteaga {\it et al}., Phys. Rev. C {\bf 79}, 034311 (2009).



\bibitem{yos08}
K.~Yoshida and N.~V.~Giai, Phys. Rev. C {\bf 78}, 064316 (2008).

\bibitem{ter05}
J.~Terasaki {\it et al}, 
Phys. Rev. C {\bf 71}, 034310 (2005).


\bibitem{dob84}
J.~Dobaczewski {\it et al}., Nucl. Phys. {\bf A422}, 103 (1984).

\bibitem{per08}
S.~P\'eru and H.~Goutte, Phys. Rev. C {\bf 77}, 044313 (2008).

\bibitem{los10}
C.~Losa {\it et al}., Phys. Rev. C {\bf 81}, 064307 (2010).

\bibitem{ter10}
J.~Terasaki and J.~Engel, arXiv:1006.0010.

\bibitem{bar82}
J.~Bartel {\it et al}., Nucl. Phys. {\bf A386}, 79 (1982).


\bibitem{yam09}
M.~Yamagami {\it et al}., Phys. Rev. C {\bf 80}, 064301 (2009).









\bibitem{ENSDF}
http://ie.lbl.gov/ensdf/



\bibitem{cha97}
E.~Chabanat {\it et al.}, 
Nucl. Phys. {\bf A627}, 710 (1997).

\bibitem{zil02}
A.~Zilges {\it et al.}, 
Phys. Lett. B {\bf 542}, 43 (2002).

\bibitem{paa03}
N.~Paar {\it et al.}, Phys. Rev. C {\bf 67}, 034312 (2003).

\bibitem{sar04}
D.~Sarchi {\it et al}., Phys. Lett. B {\bf 601}, 27 (2004).

\bibitem{lit10}
E.~Litvinova {\it et al}., Phys. Rev. Lett. {\bf 105}, 022502 (2010).


\bibitem{sood91}
P.~C.~Sood, D.~M.~Headly and R.~K.~Sheline,
At. Data Nucl. Data Tables {\bf 47}, 89 (1991).

\bibitem{PR08}
D.~Pe\~na Arteaga and P.~Ring, Phys. Rev. C {\bf 77}, 034317 (2008).
\end{thebibliography}
\end{document}